# In Search of Lost QoS

Kalevi Kilkki & Benjamin Finley

Aalto University
Espoo, Finland

## Abstract

The area of quality of service (QoS) in communications networks has been the target of research for already several decades with tens of thousands of published journal and conference papers. However, the practical introduction of QoS systems in commercial networks has been limited (with a preference for simple overprovisioning). Despite this dissonance, most influential QoS papers do not discuss this lack of penetration or challenge any of the common assumptions used to argue for QoS systems. So far, the few critical QoS papers have had only a minor effect on QoS research and standardization. Therefore, there is a serious risk that QoS will remain an academic research topic without significant practical relevance.

To help elucidate these issues, in this work, we first perform a comprehensive review of QoS including a general overview and an analysis of both influential and critical work from the past 30 years. We examine properties such as citations, keywords, and author traits to show that QoS has passed through several distinct phases with different topics while maintaining the overall attitude towards the role and objective of QoS systems. We then discuss QoS as a social phenomenon and in the context of current networking standards. Finally, we propose a QoS scheme based on incentives that avoids some of the problems identified in critical work, and we provide simple recommendations for network operators. Overall, we hope to spark the community to take a fresh look at QoS.

## Introduction

Modern communications networks form an extremely complex edifice. Information units are transmitted through fibers and air with an immense speed between network nodes making billions of decisions every second to steer each unit in the right direction. Most of the time, no human needs to bother with what happens inside the network. Sometimes information does not reach the correct destination regardless of the best effort of the network equipment. If information is lost too often or during critical moments, the quality of the applications used by the customers is deteriorated, which leads to dissatisfaction, complaints, and turnover of customers. During the last forty years, numerous sophisticated Quality of Service (QoS) mechanisms have been developed to prevent the undesirable consequences of imperfect networks.

The logic behind the use of QoS mechanisms seems to be irrefutable. Moreover, the word quality has a strong positive connotation, which strengthens the attractiveness of QoS. Therefore, many developers of communication networks share a common opinion that a network with clever QoS capabilities is better than a network without them. However, regardless of the obvious need for QoS and intense work to develop the necessary methods and mechanisms, the deployment of most QoS mechanisms have remained limited or non-existent in real networks. It seems that the true nature of QoS has been lost somewhere in the complex reality of the communications ecosystem.

This paper aims to find such a role for QoS that serves the needs of all important actors in the communications ecosystem. Because of the long and productive history of QoS, we need to build our reasoning on a strong basis. First, we present a technical framework that consists of five (mostly technical) levels that also represent different mindsets related to the nature of communications services. Secondly, we outline how the topics and solutions in the field of QoS research have evolved during the last 30 years. Thirdly, we claim that QoS research is a social phenomenon that has led to suboptimal solutions both in the focus of QoS research and in the standardization of QoS mechanisms. Finally, we propose a solution that avoids the most serious problems of the current QoS approaches.

We invite feedback to this first version of the paper. Based on the feedback, we will prepare a second version of the paper in which we explicitly acknowledge all relevant comments and proposals. We also note that we provide a list of abbreviations in an appendix at the end of the paper.

## Technical framework

QoS can be viewed from many angles. From an engineering viewpoint, the task of the network is to transmit basic information units or bits between end points quickly and correctly. The core service of a network operator is, thus, a large set of bit pipes available for the users of the network. However, not all operators appreciate that kind of limited role (Cuevas et al. 2006). In order to broaden their role, most operators want to function on the higher layers of value creation up to the level of customers and business models.

### The five levels of QoS

When the operator attempts to broaden its role, it must build on the basis of bit transmission but at the same time consider the levels above bits. The first level above bits is the level of packets or some other similar information unit. The meta-data (i.e., header) available in the packets, when properly designed, makes it possible to apply different treatments: some packets are served immediately while others are put in a queue, or even discarded. This is the lowest technical layer that allows practical QoS decisions.

However, few users are interested in the fate of individual packets; they just want their expectations to be satisfied when they are doing something requiring transmission of information. Typically, the need for transmitting information is much larger than what an individual packet can contain; as a result, consecutive packets form flows that can be continuous (e.g., voice call) or sporadic (e.g., web browsing). The quality perceived by users then depends on the flow level characteristics rather than on the fate of any individual packet. Thus, there seems to be a good justification that QoS should be defined, measured and handled on the level of flows. However, there can be millions of concurrent flows in a large network which makes it is challenging to handle each flow individually. Thus, many network operators want to combine flows into aggregate level units that supposedly are easier to manage and control. Aggregation is typically based on the similarity in quality requirements or customer contracts.

In IP technology, the distinction between connection (or flow) level and aggregate level is pronounced due to historical reasons: aggregate level DiffServ was developed as a solution to the problems of the flow level IntServ (see, e.g., Gozdecki et al. 2003). However, DiffServ uses packet level marking and mechanisms, which obscures the boundaries between the levels.

In reality, it is often difficult to distinguish separate flows and their specific needs, which makes the construction of feasible aggregates difficult. In the case of an access network, identifiable units include devices and subscribers in addition to flows. Without a robust identification, the outcome of any QoS action remains unsure and vague. Thus, QoS actions are often directed to clearly identified entities: devices in the case of WLAN network and subscribers in the case of cellular networks.

Thus, quality of service can refer to different levels from transmission of packets to user satisfaction. Because packets, flows, and aggregates do not feel anything, the quality on those layers has to mean an objectively measurable property. In contrast, as the human mind is difficult to measure objectively, quality has to refer to the subjective assessment of satisfaction with the service or the usefulness of the service. As a conclusion, we use five viewpoints called packet, connection, aggregate, device, and subscriber as summarized in Table 1.

Table 1. QoS from five viewpoints.

| Level or viewpoint | Main objective and QoS criteria | Technology & marking | Typical pricing | Trust on end device | Related terms |
|---|---|---|---|---|---|
| Subscriber | Service experience compared to expectations (QoE) | Cellular (SIMA) | Flat rate, (or per MB) | Based on SIM card | Customer, user, consumer |
| Device | Connectivity (in general) | WLAN | Based on access time | Medium | Station, application, vendor |
| Aggregate | SLA (operational efficiency) | ATM VP IP DiffServ MPLS 5G slice | Based on SLA | Medium | Path, bearer, class, slice, route, stream, traffic, VPN |
| Connection | Ability to meet pre-defined quality requirements (Call blocking) | ATM VC IP IntServ | Based on connection parameters | Low | Call, flow, session, circuit |
| Packet | Delay, delay variation, packet loss ratio | IP TOS field ATM cell loss priority | Flat rate | High | Protocol data unit, (bit, byte) |

As Table 1 presents, the terminology is diverse on all of the levels. In this framework, a flow is a continuous series of packets without exact predefined properties whereas a connection is controlled and managed path through a network. Day (2008, p. 28) has presented a similar distinction between flow and connection based on the level of shared state inside the network. Technically, it is possible to define an IP flow based on the IP header information: packets with the same source and destination address, protocol field, source port, and destination port belong to the same flow. Many flows (or applications that generate the flows) are often adaptable to variable network conditions, while connections usually are not.

Many technologies can be located on the aggregate level. One them is Virtual Private Network (VPN) using public Internet as the transmission medium. VPNs need isolation that is often realized by using another aggregate level technology, Multiprotocol Label Switching (MPLS). These are tools for big network operators. As an example on another end of the scale is the management tools in the Linux-based firmware for wireless routers, DD-WRT. DD-WRT (2018) gives the possibility to map traffic into five classes (maximum, premium, express, standard, and bulk) based on application, interface, IP address, MAC address, and port number. The operator can give different shares of capacity for each of the classes. The necessary actions are easy to perform without any network engineering background. The apparent simplicity of this QoS method hides the fact that it is very hard for even a competent professional to deduce how the selected shares affect user experience.

From the perspective of networking technology, an application is an interface between a user and a network and, thus, an integral part of almost every QoS analysis. Users, customers, and subscribers are similar terms. Users can remain unidentified and somewhat obscure for network equipment whereas subscribers make contracts that typically last a limited period. Thus, subscribers are identifiable though not necessarily as individual persons. A customer is a person or another legal entity that pays for the service used by one or several users.

QoS actions

So far, we have used general term of QoS to describe the overall phenomenon related to any quality aspect of communication networks and services. To deepen our analysis, we need to consider more thoroughly what happens in real networks. We use the term QoS mechanism to refer to the concrete actions inside the network that affect the QoS on any of the five layers. In principle, there are five basic types of actions: serve instantly, postpone, change priority, limit, and reject. The content of these actions on different layers are explained in Table 2.

Table 2. Possible QoS actions.

| Action Level | Serve immediately | Serve later | Change the status (marking) | Limit the size | Reject |
|---|---|---|---|---|---|
| Subscriber | Accept SIM | - | Change service class | Change access rate or data volume | Deny access right |
| Device | Connect device | - | Change priority | Change access rate | Disconnect device |
| Aggregate | Establish a path | Pre-schedule a path | Change priority | Change maximum rate | Terminate a path |
| Connection | Accept a connection | Put a connection in waiting queue | Change priority | Change maximum rate | Reject a connection attempt |
| Packet | Transmit packet | Put packet in a queue | Change type of service marking | - | Drop packet |

All QoS actions should take into account the available resources and when relevant information is available, the importance or urgency of the information unit. In addition, some actions are difficult to locate on any of the levels shown in Table 2. The operator may also change the overall QoS scheme that defines how different types of QoS mechanisms are used to meet the business objectives of the operator. Those actions need to be converted to actions on the levels shown in Table 2.

Another important matter affecting the outcome of a QoS action is how end devices and users react to the properties of ensuring network service. The three main types of feedback mechanism are 1) the flow control implemented in Transfer Control Protocol (TCP), 2) adaptive video or audio coding, and 3) user reactions, e.g., the user ends the session due to low quality or changes the operator. The standpoint of this paper is that these reactions are not included in QoS mechanisms because they are not in direct control of the network. Nevertheless, they considerably affect the relationship between QoS actions, the measurable network performance, and user perception, and, hence, must the included in the overall analysis of service quality (see De Cicco et al. 2011 as an example of adaptive video). Similarly, we do not consider pricing as a QoS method although it is closely related to QoS and it can be used to solve similar problems as QoS mechanisms do, in particular during overload situations (see, e.g., DaSilva 2000).

Another key concept is the application. In the context of QoS, the application is a computer program that provides a user with tools to accomplish a task, for instance, making a phone call, browsing in the web, or playing a game. Different applications (and their implementations) have different requirements for network services. Those requirements are often presented as the main justification for the use of QoS mechanisms as discussed later in this paper. Application requirements concern both packet and flow level properties. This mindset is illustrated in Figure 1. Note particularly that the process flows from left to right without any apparent feedback loop as both the user and the network make their decisions before the connection is established.

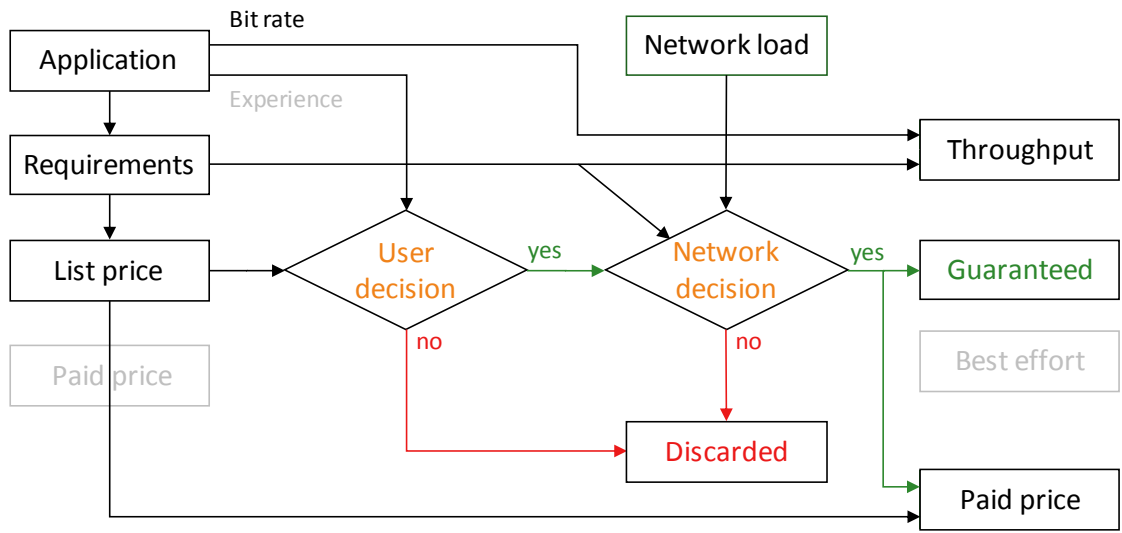

Figure 1. An outline of the service process based on connection-oriented QoS.

The connection-oriented model can be contrasted with the simplest packet-oriented model, that is, the best-effort model used in the Internet, illustrated in Figure 2. In this model, the application does not explicitly require anything. However, there is an inherent feedback loop from the momentary throughput (or packet loss ration) back to the application. The application may react to the situation by changing the sending bit rate. Users make decisions about whether to continue with the application based on their momentary experience. These two models represent the two extremes in the field of QoS: guaranteed quality based on application requirements, and best effort model without any guarantees or QoS. Various QoS methods between these two extremes have been proposed during the last thirty years as the following section shows.

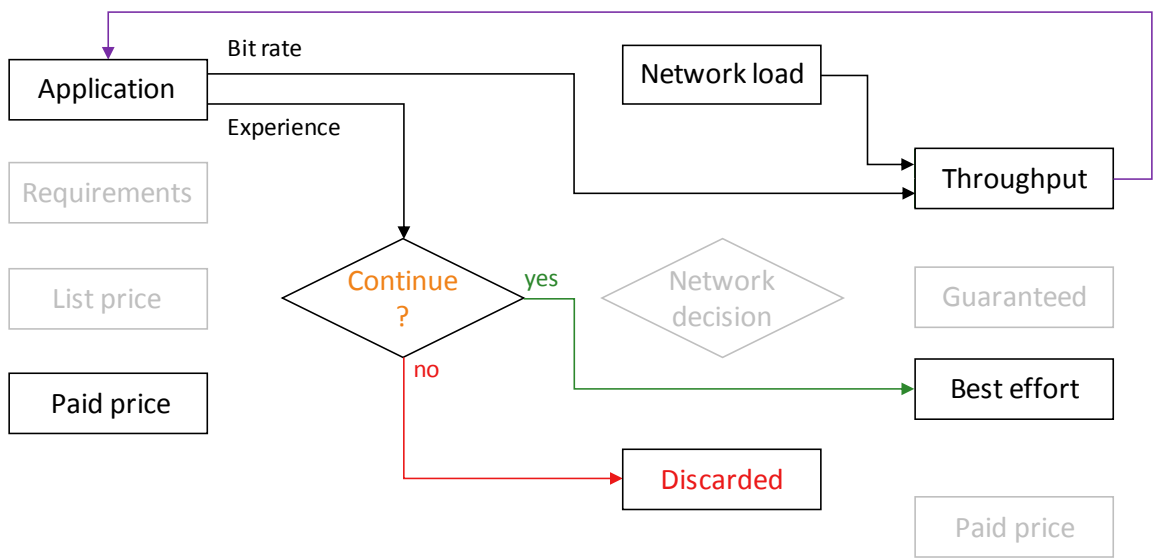

Figure 2. An outline of the service process based on packet-level, best effort transportation.

## History

This section gives a historical overview of activities that fall under the umbrella of QoS. Our analysis is based on a literature study that covers both the most cited and critical QoS publications. In addition, the literature review includes a set of papers that have contributed to the development of new viewpoints of QoS. The main material is a collection of 270 QoS papers from the IEEE Xplore database. The selection process is as follows 1) all papers with QoS in the abstract are selected, 2) papers with a non-networking topic are removed (mostly dealing with cloud or Web services), 3) the papers for every year from 1991 to 2017 are ordered based on the number of citations, 4) the ten most cited papers in each year are selected to the "most cited papers" data set.

As to the history of QoS, in addition to longer discussions in Willis (2005), Xiao (2008), and claffy & Clark (2016), several papers provide brief QoS histories in the context of the Internet, see e.g., Carpenter & Nichols (2002), Crowcroft et al. (2003), Parekh (2003), Hutchison (2008), Schulzrinne (2010), and BITAG (2015). Hiertz (2010) discusses the history of QoS in the context of IEEE.

The number of published QoS papers showed rapid growth from 1990 to 2008 as presented in Figure 3. The growth turned down around 2009 even if we take into account that the concept of Quality of Experience (QoE) partly replaced QoS as the key term in some studies. As to QoE, de Souza & Dantas (2015) have presented a similar figure with data gathered from Web of Knowledge. Figure 3 also shows the number of patents with QoS in the abstract in the US patent office database. The number of QoS-related patent applications has been relatively stable after 2005.

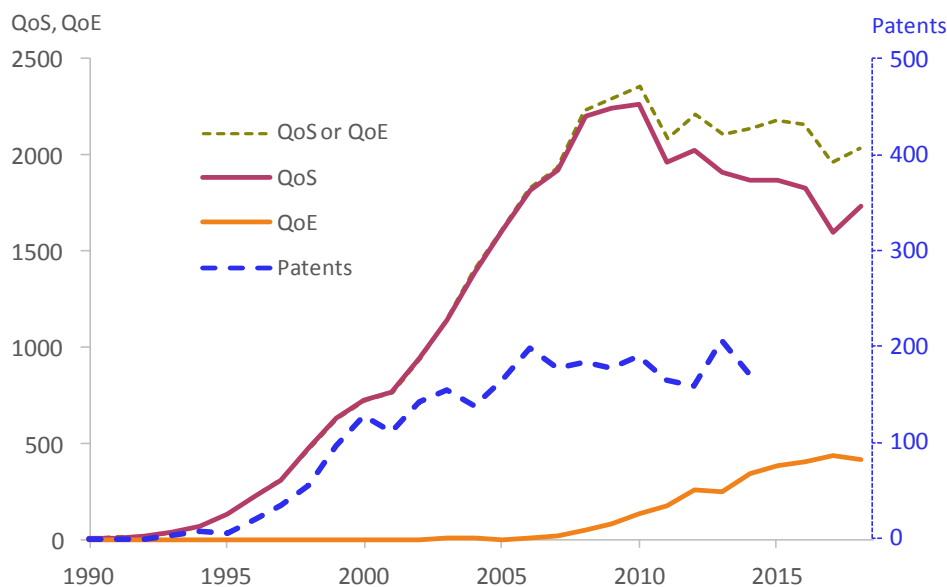

Figure 3. Papers per year with QoS, QoE, and "QoS or QoE" in the abstract in IEEE Xplore database. The number of patents figures (right scale) are collected from the US patent office database.

Table 3. Topics in the most cited QoS papers 1991 - 2017.

| Year | ATM | IP | Cellular | WLAN | Ad-hoc | QoS routing | Cognitive radio | QoE | Other |
|---|---|---|---|---|---|---|---|---|---|
| 1991 | 8,0 | | | | | | | | 2,0 |
| 1992 | 9,0 | | | | | | | | 1,0 |
| 1993 | 9,0 | | | | | | | | 1,0 |
| 1994 | 9,0 | | 1,0 | | | | | | |
| 1995 | 5,0 | | 2,0 | | 1,0 | 1,0 | | | 1,0 |
| 1996 | 4,5 | | 4,0 | 0,5 | | 1,0 | | | |
| 1997 | 2,5 | 1,5 | 3,0 | | | 3,0 | | | |
| 1998 | | 1,0 | 4,0 | | | 3,0 | | | 2,0 |
| 1999 | | 3,0 | 1,0 | | 3,0 | 3,0 | | | |
| 2000 | | 7,0 | | | | 1,0 | | | 2,0 |
| 2001 | | 1,0 | 5,0 | | 1,0 | 3,0 | | | |
| 2002 | | 2,0 | 2,5 | 0,5 | 1,0 | 2,0 | | | 2,0 |
| 2003 | | 1,0 | 1,5 | 5,5 | 1,0 | 1,0 | | | |
| 2004 | | | 0,5 | 4,5 | | 3,0 | | | 2,0 |
| 2005 | | | 2,5 | 4,5 | 1,5 | 0,5 | | | 1,0 |
| 2006 | | | 2,0 | 6,0 | | 2,0 | | | |
| 2007 | | | 2,0 | 5,0 | 1,0 | | 2,0 | | |
| 2008 | | | 2,5 | 1,5 | | | 5,0 | | 1,0 |
| 2009 | | | 2,5 | 6,5 | 1,0 | | | | |
| 2010 | | | 1,5 | 3,5 | | | 3,0 | 2,0 | |
| 2011 | | | 2,5 | 2,5 | 1,0 | | 2,0 | 1,0 | 1,0 |
| 2012 | | | 4,0 | 3,5 | | 1,5 | 1,0 | | |
| 2013 | | | 6,0 | 1,0 | 0,5 | 1,5 | 1,0 | | |
| 2014 | | 1,0 | 6,0 | 2,5 | | 0,5 | | | |
| 2015 | | | 6,5 | 0,5 | 1,0 | | 1,0 | 1,0 | |
| 2016 | | | 7,0 | | 1,0 | | 2,0 | | |
| 2017 | | | 6,5 | 1,5 | | 2,0 | | | |

Table 3 shows the main topics in the most cited QoS papers. The papers are classified primarily based on the target network technology (ATM, IP, etc.) and secondarily on the specific topic (e.g., QoS routing). In several cases, two topics have selected either because the paper covers several technologies (e.g., cellular and WLAN) or the paper deals with certain aspect (e.g., routing) only in the context of a specific technology (e.g., ad-hoc networks).

We can divide the history of QoS into five phases mainly based on the dominant QoS topic in the set of most cited papers shown in Table 3 as follows:

- Prehistory, until 1990
- ATM, 1991 – 1996
- Diversity, 1997 – 2002
- Wireless rules, 2003 – 2009
- Back to cellular, 2010 – 2017

The boundary between ATM and Diversity periods coincide the watershed between old and new telecommunications paradigms with different terminologies (e.g., from monopoly to competition and from cable to mobile) as analyzed by Kwon & Kwon (2017).

In addition to the topics of the QoS papers, we also examined the terminology used in the papers. We first pick the first three sentences that contain acronym QoS. Table 4 presents the share of papers in which a term (e.g., user or application) appears at least once in those three sentences. According to this data shown in Table 4, the changes in the vocabulary has been relatively small. Most notably, guarantee, traffic, and connection were more popular during the ATM period, while user and experience have been more popular during the last period. The changes in the frequency requirements have been minor, which is an indication of the stability of general attitude towards QoS. Some important concepts are conspicuously rare in the 790 sentences: customer is mentioned three times, TCP, subscriber, and business twice, and profit once. Terms that do not appear in the sentences include revenue, churn, feeling, happiness, and frustration.

While the share of QoS papers without mathematical formulas has remained relatively stable, the share of QoS papers with more than 20 formulas has increased during the last fifteen years.

Table 4. Authors, terms, and formulas in the set of the most cited QoS papers.

|  |  | ATM 1991-96 | Diversity 1997-2002 | WLAN rules 2003-09 | Back to cellular 2010-17 | All periods |
|---|---|---|---|---|---|---|
| Bachelor degree | North America | 26 % | 15 % | 9 % | 3 % | 12 % |
|  | Europe | 24 % | 29 % | 20 % | 11 % | 20 % |
|  | Asia | 47 % | 49 % | 66 % | 81 % | 62 % |
| PhD | North America | 50 % | 75 % | 63 % | 28 % | 52 % |
|  | Europe | 28 % | 18 % | 24 % | 23 % | 23 % |
|  | Asia | 18 % | 7 % | 11 % | 48 % | 23 % |
| Highest degree | Electrical engineering | 66 % | 53 % | 50 % | 49 % | 54 % |
|  | Computer science | 19 % | 37 % | 41 % | 18 % | 28 % |
|  | Non-engineering | 7 % | 6 % | 1 % | 0 % | 3 % |
| Employer | North America | 60 % | 82 % | 63 % | 26 % | 56 % |
|  | Europe | 25 % | 12 % | 21 % | 25 % | 21 % |
|  | Asia | 13 % | 7 % | 14 % | 48 % | 22 % |
| Type of employer | University | 58 % | 67 % | 83 % | 85 % | 74 % |
|  | Operator | 18 % | 3 % | 0 % | 1 % | 5 % |
|  | Vendor | 20 % | 27 % | 14 % | 6 % | 16 % |
| Gender | Female | 7 % | 0 % | 16 % | 18 % | 11 % |
| Published in | Journal | 47 % | 82 % | 93 % | 96 % | 81 % |
| Terms used in the first 3 sentences with QoS | application | 17 % | 32 % | 33 % | 21 % | 26 % |
|  | connection | 25 % | 7 % | 4 % | 4 % | 9 % |
|  | experience | 2 % | 2 % | 1 % | 13 % | 5 % |
|  | guarantee | 40 % | 28 % | 21 % | 29 % | 29 % |
|  | requirement | 55 % | 43 % | 41 % | 54 % | 49 % |
|  | traffic | 52 % | 27 % | 26 % | 16 % | 29 % |
|  | user | 32 % | 30 % | 37 % | 46 % | 37 % |
| Number of formulas | 0 | 38 % | 35 % | 20 % | 28 % | 30 % |
|  | > 20 | 13 % | 15 % | 31 % | 43 % | 27 % |

As to the most popular QoS topics, a transition from Wi-Fi back to cellular technologies started around 2010. At the same time, the center of gravity of QoS research started to move from North America to Asia as illustrated by the PhD and Employer rows in Table 4. Other observations from Table 4 include the prevalence of engineering education, mostly in the area of electrical engineering and computer science. Non-engineering education is rare: there are six first authors with degrees in mathematics, only two in economics and one in sociology among the 270 papers.

Compared to the papers published in Telecommunications Policy (Gómez-Barroso et al. 2017), the share of North America is somewhat larger in the case of QoS (56 %) than in the case of policy issues vs. (41 %), when the country is defined by the affiliation of the authors, and the share of Policy papers refers to the share of citations. The share of Asia is slightly larger in the case of QoS (22 %) than in the case of telecommunications policy (16 %).

The share of female authors in QoS papers has increased over time but has remained below 20 %. A great majority of authors worked in a university whereas the share of authors working at a network operator has been negligible after ATM period. Most of the papers have been published in journals. The most popular journals have been IEEE Selected Areas in Communications (45 papers), IEEE Communications Magazine (29 papers), and IEEE Surveys and Tutorials (19 papers). The most common conference is Globecom (19 papers). Altogether, the 270 most cited QoS papers have published in 36 different journals and 12 different conferences or seminars. As to citations, the set of 270 most cited papers represents about 1 % of all QoS papers in IEEE Xplore database, but they have gathered roughly 25 % of all citations to all QoS papers.

The other main material for the QoS history is a set of 13 papers and presentations offering a credible analysis of the problems of QoS containing Bell (2003), Bhagat (2006), Bricklin (2003), Corbató & Teitelbaum (2006), Huston (2012), Hutchison (2008), Isenberg (1998), Mackay & Edwards (2013), Meddeb (2010), Odlyzko (1998), Pras et al. (2005), Schulzrinne (2010), and Turner (2007). Because the set is small and the selection is based on a subjective assessment, it does not allow any formal statistical analysis. We can, however, note that in most of the critical papers and presentations there is only a single author while in the most cited papers the share of single-author papers is below 6 %. Only one of the critical papers was published in a traditional journal. The average number of citations to the critical papers per Google Scholar is 30, which is, of course, much lower than in the set of most cited papers, but it is still roughly twice as high as the average number of citations for all QoS papers.

Prehistory until 1990

The word *quality* comes from Latin *qualitatem* coined by Cicero to translate Greek word *poiotes* ("of-what-sortness") which, in turn, was possibly coined by Plato (Online Etymology Dictionary 2018), O'Conner & Kellerman 2012). This intricate history may partly explain the elusiveness of quality. Service comes from Latin servitium "slavery, condition of a slave, servitude." Thus, in ancient Rome, quality of service referred to the characteristics of slaves.

Telephone service was one of the first contexts in which the quality of the service provided by different companies was systematically assessed. Valentine (1911) lists as essential qualities of telephone service accuracy, speed of connection, and uniformity in speed and methods of handling calls. Systematic methods were developed to inspect telephone services based on those quality factors. However, before 1982 the term *quality of service* was used only occasionally in the context of telephone or any other communications service.

In the 1980s, three processes affected how quality of service was perceived: rapid development of information technology, deregulation of communication business, and research activities in management science. The digitalization of information technology made it easier for new entrants to compete with old players both in the vendor and operator business. This process led gradually to an increased diversity of service offerings—but only when regulators allowed or even promoted competition. However, it is hard to assess whether improving service quality was a significant motive for the deregulation of telecommunications services. It was even anticipated that the quality of service perceived by basic telephone users could decline due to deregulation (Horwitz 1991, p. 274). In any case, after a long period of monopolies, competing service offerings created a need to consider quality of service from novel perspectives.

At the same time, the relatively small and restricted ARPANET developed into the large-scale, open and global Internet we are accustomed to today. The mindset when dealing with problems of controlling traffic load was limited to well-defined IP packets and the somewhat obscure concept of flow without much reflection on quality of service. Nevertheless, the IP packet (or datagram) header included the Type of Service (TOS) field with three precedence bits and one bit for delay, throughput, and reliability (Postel 1981a, RFC 791). Precedence bits made it possible to classify packets into routine, priority, immediate, flash, flash override, emergency, and network control and internetwork control. The naming and logic of precedence bits originated from military messaging systems (Baran 1964, Bauer et al. 2009, and claffy & Clark 2016). The military had established organizations, rules, and routines that defined and controlled the use of priorities for all *messages*. In contrast, the operational staff of the Internet (working in universities and research organizations) concentrated on the packet level performance and trusted on egalitarian use of networking resources.

Furthermore, the TCP specification (Postel 1981b, RFC 795) declared that the type of service marking for TCP flows should be "routine," "normal delay," "normal throughput" and "normal reliability." There was no architectural framework to inform the usage of TOS bits (Carpenter & Nichols 2002). As a result, the precedence bits were seldom if ever used in real networks while the Internet provided one (best effort) service that delivered packets through the network without any guarantees as illustrated in Figure 2. Obviously, the mindset of IP networking in the 1980s was on the packet level mechanisms, not on any upper level.

IEEE approved the first version of 802.1D, Media Access Control (MAC) Bridges, in 1990. The standard included a QoS parameter called User priority (UP) that is the priority requested by the originating service user (IEEE 1990, 802.1D). The value of User priority ranges from 0 (lowest) through 7 (highest). The first standard did not elaborate on the use of the priorities. IEEE explained in an informative annex (IEEE 2004, 802.1D) that user priorities can be used to manage latency and throughput. In case of seven traffic queues, traffic types for different user priorities could be (from lowest to highest priority) background, best effort, excellent effort, controlled load, video, voice, and network control.

Another activity that could have affected the construction of QoS models in modern networks was the extensive literature about general product quality published in the field of management sciences. As noteworthy examples, Garvin (1984), Grönroos (1984), and Parasuraman et al. (1985) provided systematic analyses of the diverse meanings and ingredients of product quality. Garvin distinguishes five definitions of quality in the context of products. The most relevant for our purposes are User-based definition (degree of satisfaction), manufacturing-based definition (conformance to requirements), and value-based definition (usefulness for the customer). Grönroos introduced a two-dimensional model that distinguishes (1) technical quality based on service performance, and (2) functional quality based on user perception. Parasuraman et al. analyzed the four gaps in a five-part "quality chain" model: consumer expectation, management perception of the consumer expectations, service quality specification, actual service delivery, and external communications about service. Most of the QoS literature concentrates on quality specifications and actual service delivery parts in the chain while ignoring all other gaps in the chain.

At the same time, CCITT (1985) defined quality of service in Recommendation G.106 (Red Book) as follows: "Collective effect of service performances which determine the degree of satisfaction of a user of the service." This definition does not clearly state whether the primary issue is technical performance or user satisfaction, but instead vaguely refers to the process between them. Another standardization organization, ISO 8402 (1986) defined quality as "the totality of features and characteristics of a product or service that bear on its ability to satisfy stated or implied needs." As an interesting observation, both standards state that either quality of service (in G.106) or quality (in ISO8402) "is not used to express a degree of excellence in a comparative sense nor is it used in a quantitative sense for technical evaluations." CCITTs recommendation I.350 (1988) limits the use of QoS to the parameters that are directly observed and measured while subjective QoS parameters will not be specified in I-Series of Recommendations on QoS. Thus, the mindset seemed to be that QoS covers technical and subjective aspects and anything that happens between them, but those technical and subjective aspects can be defined and analyzed separately. There is some similarity here with Cartesian dualism, that is, the dualist separation of mind and body. The separation has proved to be quite permanent also in the context of QoS.

As to the analysis, design, and development of QoS mechanisms, the three tracks (Internet, management science, and telecommunications) remained more or less separate until the end of the 1980s. Then came a serious effort to design a ubiquitous broadband network.

ATM (1991 – 1996)

In the Internet, QoS did not play a significant role in the 1980s. The mindset was that any special service could be satisfied by an appropriate packet marking and by some clever protocols to be designed when needed. The mindset that directed the design of ATM was the opposite: the design shall be constructed on the basis of all identified needs of users and applications. As to traffic control, ATM has a clear and consistent philosophy with two parts (ITU-T 1993, Rec. I.371): (1) the network negotiates performance objectives for every connection and (2) the task of traffic control is to avoid situations in which the network cannot meet the agreed objectives. This approach is illustrated in Figure 1. Connection was defined as a temporary association of transmission channels and other functional units to provide the means to transfer information between endpoints. This is a perfect definition for the connection level in Tables 1 and 2. Even though ATM specifications included a Virtual Path (VP) layer to facilitate the handling of multiple connections (aggregate level) and a cell loss priority bit in the header of an ATM cell (packet level), the connection has always been the most important unit in ATM networks.

We may even state that in the context of ATM traffic was something wild that needed to be tamed and enclosed in a cage of connection. A key tool in the taming effort was mathematical modeling. For instance, the final report of COST 242 project (Roberts et al., 1996) tried to conquer broadband traffic by hundreds of mathematical formulas presented in 545 pages.

That was the main principle of ATM developers, but was it adopted among the researchers that wrote QoS papers? Definitely yes, at least in the period from 1991 to 1996. Three-quarters of the most cited QoS papers were addressing ATM directly and 40 percent of the papers included *guarantee* in the first three sentences with QoS; *requirement* was even more common than guarantee (see Table 4). None of the 60 most cited papers during this period provided any critical analysis of the basic principles of ATM. We could not find any documented linkage from general service quality literature to network QoS research during this period. The definition for QoS remained the same as in earlier recommendations (see, e.g., ITU-T 2008, T-REC E.800).

It seems that the most active researchers did not have any need to question the dominant way of thinking of QoS. Obviously, there were quality requirements that the network had to meet. Where did those magic requirements come from? If we look at the 60 most cited papers in this ATM period, we find 23 papers in which connections had quality requirements. In 12 papers, users had requirements whereas only in one paper applications had quality requirements.

Certainly, some people were skeptical about the real merits of ATM and the feasibility of guaranteed connections, but they were not actively promoting their opinions in those forums in which QoS was formally discussed. Furthermore, the results achieved in the field of management science did not appear to have any significant effect on the QoS design and analysis on the networking field. Sometimes no research at all is needed to create a valuable business. As an example, short message service appeared as a highly profitable business for mobile operators (Hillebrand et al. 2010 and Lähteenmäki et al. 2017) almost without any research or business development.

And then the Internet and Web arrived and everything changed. In the Internet there were no guarantees whatsoever; still the Internet provided services that users, organizations, and companies were immediately eager to utilize while ATM waited for practical applications and a feasible business model. Even though some notable IETF experts raised the question whether the Internet also requires special services for some demanding applications, (see Braden et al. 1994, RFC 1633), the tremendous Internet growth in the 1990s proved the feasibility of the simple best-effort service.

### Diversity (1997 – 2002)

The nature of QoS papers started to change around 1996. It became gradually apparent that the connection-oriented model used in ATM networks was not justified economically. Guaranteed quality was an insufficient reason for ordinary consumers to pay any extra fee compared to the ubiquitous Internet access based on a monthly fee. Finally, ATM was essentially abandoned as a QoS topic. Instead, the main QoS topics during this period of diversity were IP, cellular, and wireless networks as shown in Table 3.

As to IP networks, an informational architecture document for Integrated Services (IntServ) was published already in the heydays of ATM (Braden et al. 1994, RFC 1633), while other key specifications, e.g., RFC 2205 (Braden et al. 1997), RFC 2210 (Wroclawski 1997), and RFC 2212 (Shenker et al. 1997), were published when ATM's reputation was already in decline. The second active standardization track with an IntServ-like mindset was the specification of QoS routing by Crawley et al. (1998, RFC 2386). They defined QoS as "a set of service requirements to be met by the network while transporting a flow." The overall mindset in both the IntServ and QoS routing context was a kind of mixture of a connection-oriented process (Figure 1) and best effort process (Figure 2). The third QoS track in IETF was Differentiated Services (DiffServ) that resulted in key specifications in Blake et al. (1998, RFC 2475). The motivation for DiffServ was operational efficiency that necessitated the handling of aggregate traffic streams. Another networking technology on the aggregate level, Multiprotocol Label Switching (MPLS), was specified by Rosen et al. (2001, RFC 3031). In reality, even after these specifications were implemented in network equipment, the Internet still worked based on best-effort model (packet level) while operational burden was somewhat alleviated by means of aggregate-level protocols and mechanisms, that is, by MPLS and DiffServ.

In cellular networks, the first data service was based on circuit switching that allowed 42 kbit/s throughput. However, it was already then obvious that circuit switching was not a reasonable solution for web-traffic due to the diversity and variability of data applications (Hillebrand 2013 and Bettstetter et al. 1999). Thus, General Packet Radio Service (GPRS) was developed to offer flexible data services over GSM networks. ETSI (1999) defined GPRS QoS profile that consists of four parameters: service preference (priority), reliability, delay, and user data throughput. According to the specification, reliability is requested by an application while throughput is requested by the user. In contrast, the specification does not indicate who requests a specific priority or delay class.

The next generation of cellular technology, UMTS (later called 3G), defined four QoS traffic classes: conversational, streaming, interactive, and background. The term *class* seems to refer to aggregate level and a DiffServ type of construction. The conversational class offers a strictly connection-oriented service while the background class offers a best effort service that is packet-oriented. Nevertheless, the main difference between the UMTS classes is in delay constraints. Also, the UMTS specification defined traffic handling priority, but only for the interactive class.

Although network technologies with and without QoS evolved, the changes in the general mindset were modest if measured by the vocabulary in the first three QoS sentences in the most cited QoS papers (see

Table 4). The vocabulary used in QoS papers was somewhat adjusted, but the mental model based on quality requirements was persistent. However, a part of QoS community became aware that a majority of QoS mechanisms remained unused in real networks. As a result, some critical papers started to emerge at the end of the decade. Isenberg (1997) had a simple message for network operators: "Just deliver the bits, stupid." Despite the popularity of the Stupid Network meme, none of the 270 most cited QoS papers refer to Isenberg (1997).

The series of IWQoS workshops was the main forum for discussing the benefits and challenges of QoS. For instance, IWQoS '97 included a panel discussion about the need for reservations in IP networks (see Baker et al. 1997). The common opinion among the top experts was that some applications need special QoS, whereas a strict reservation (using RSVP) was not necessarily the best solution in practice. A year later in the same series of workshop Guerin (1998) was concerned about the cost, need, and efficacy of QoS provision in his keynote address. However, because QoS mechanisms were still under development, most experts remained more optimistic about the need and use of QoS in the internet. For instance, Internet2 consortium of US research and education institutions made a comprehensive plan, named QBone, to use QoS mechanisms to support real-time applications (Teitelbaum & Shalunov 2008). However, QBone was never used in any operational environment.

While the majority of QoS papers were technical, a few publications assessed QoS from non-technical perspectives. Ang & Nadarajan (1997) noted that QoS would be a key issue in the regulation of Internet services. The main problems they noticed were the overly technical nature of QoS standards and the Internet design principle based on inherently imperfect service. Odlyzko (1998) took an economic perspective and concluded that over-provisioning is the most reasonable approach and when unfeasible, the simplest possible QoS scheme should be used to minimize the burden on the users and network operators.

Gray (1999) divided people into three groups based on their opinion about QoS: optimists, pessimists, and fence-sitters. In Gray's terminology, optimists believed the cost of network capacity would become so low that no complex mechanisms are needed while pessimists believed that bandwidth is always scarce. Fence-sitters wanted to be optimists but thought that it would be necessary to have a contingency plan in case pessimists were right. The strategy promoted by Gray was to support multiple traffic classes but without per-flow admission control. Similarly, Xiao & Ni (1999) noted that the need for QoS is a hotly debated issue: some believed that optical technology would make transmission capacity so abundant that there would be no need for QoS while others believed that new applications would always consume all available resources and cause congestion.

Ferguson & Huston (1998, p. xvii) argued that marketing personnel had high expectations about the additional revenue generated by QoS, while network engineers wanted to provide predictable behavior for a particular application. A couple of years later, Huston (2000a, 2000b) concluded that the combination of IntServ and DiffServ was not sufficient to allow wide-scale deployment of QoS in the Internet, because of two reasons. First, Internet-wide QoS would need complex multi-party co-ordination and financial settlements between operators, and secondly, the availability of very high capacity transmission systems diminish the need for QoS. Pan & Schulzrinne (2001) stated that though overprovisioning could be a feasible solution in some backbone networks, it is not a cost-effective solution for all ISPs.

Thus, some critical questions about QoS occasionally emerged, but the discussions remained limited in the sense that it did not affect considerably on the mainstream QoS studies and on the QoS standardization. In Gray's terminology, most of the critical authors, except Isenberg, were fence-sitters, while the mainstream QoS studies were based on a pessimistic attitude (that is, network resources always remain scarce). Nevertheless, the discussions led to a new field of QoE studies that emphasized the importance of understanding users' needs and experience, see, e.g., Bouch et al. (2000), van Moorsel (2001), and Khirman & Henriksen (2002) as early contributions in this field. QoE is similar to the functional quality based on user perception defined by Grönroos (1984), although hardly any QoE paper refers to Grönroos. In general, the first network QoS paper that referred to Grönroos (1984), Garvin (1984), or Parasuraman (1985) was Ludwick & Grant (2002)—thus a delay of 18 years.

Wireless rules (2003 – 2009)

The first significant event in which QoS was discussed extensively was the ACM SIGCOMM workshop on "Revisiting IP QoS: What have we learned, why do we care?" (RIPQOS) held in August 2003. The workshop attracted not only scientists and researchers but also people with practical experience with the deployment of QoS. The overall result of the presentations and discussions were outlined by the workshop Chair Grenville Armitage (2003).

Bell (2003) argued that the main reason for the failure of QoS has been the missing feedback loop between network operations and QoS research. This missing feedback led to the excessive complexity of QoS systems and the limited use of QoS in real networks. Teitelbaum & Shalunov (2003) stated that QoS is needed only in worst-case situations, e.g., when the network is suffering from Denial of Service attacks. Henderson & Bhatti (2003) studied the need for QoS in the context of networked games. They noticed that although delay affected game performance, players are unlikely to pay for higher levels of QoS. Davie (2003) remained relatively optimistic about the use of QoS in the future while stressing that most of the barriers to wider deployment are business related rather technology related. Crowcroft et al. (2003) stated that QoS provides network operators a means to improve their business by charging for different levels of service. Finally, Burgstahler et al. (2003) represented the conventional viewpoint in which QoS is needed because of the substantial diversity in the requirements set by different users and applications. Thus, there was no common agreement about the need for QoS mechanisms. As far as we know, no similar discussion forum has been arranged after the RIPQOS workshop.

In addition to RIPQOS papers, Bricklin (2003) argued that in packet networks it is much easier and more efficient to add capacity than to build elaborate QoS mechanisms providing only minor improvement during congestion. Parekh (2003) noticed that QoS systems were not used and gave as possible reasons the lack of a business model, lack of congestion, billing problems, and that popular applications do not need QoS. Regardless of the merits of the critical papers, the discussion about the feasibility of QoS did not reach the general audience.

The focus of QoS research clearly moved to wireless networks around 2003 as shown in Table 3. About half of the top-10 QoS papers during the period from 2003 to 2009 analyzed QoS in wireless networks (IEEE 2005, 802.11) while only a couple top-10 papers were related to IP networks. Also, WiMAX (IEEE 2009, 802.16) gained considerable popularity among QoS researchers at the end of the period and at the beginning of the next period. The basic unit for QoS control in WiMAX is a unidirectional service flow with numerous traffic parameters. In general, the traffic control in WiMAX resembles that of ATM with an additional priority parameter on the level of service flows. In addition to conventional wireless and cellular technologies, cognitive radio was popular for a brief period in 2007 - 2008.

The papers and presentations in RIPQOS seem to have had no effect on the development of QoS wireless networks. A typical assumption was that QoS is needed in radio access because bandwidth is always limited and requires special attention, even when it is assumed that the huge capacity of optical fibers is large enough to justify the use of the best effort model in the fixed Internet.

The standardization of QoS mechanisms in IEEE 802.11 networks was based on a belief that some multimedia applications require mechanisms that give QoS guarantees over a wireless link (see, e.g., two papers, Mangold et al. 2003 and Zhu et al. 2004, that belong to the set of most cited QoS papers). The QoS amendment for IEEE 802.11 was published in 2005 (IEEE 2005). IEEE uses the term access category as a label for a set of channel access parameters that are used to transmit data units with pre-defined priorities. The IEEE standard defines four access categories: best effort, background, video, and voice. Background is a less than best-effort service while video and voice categories have delay limits, with voice somewhat stricter than video. Due to the nature of wireless (WLAN) networks, the main viewpoint is device-oriented. Yet, access categories are partly based on applications mainly because of the different delay requirements. In practice, high-priority stations are polled more frequently, and they get more transmission capacity than low-priority stations.

Thus, by 2005, all major standardization organizations had defined QoS mechanisms based on aggregate level logic. In 1999, IETF defined four Assured Forwarding (AF) classes (Heinanen et al., 1999), then in 2000, 3GPP defined four QoS classes (ETSI 2000) and finally, in 2005, IEEE defined four access categories (IEEE 2005). In addition, all of these standards include another dimension called drop priority, traffic handling priority, or user priority depending on the standardization organization.

As to the most critical QoS analysis, the main papers during this period were Pras et al. (2005), Turner (2007), Corbató & Teitelbaum (2006), and Bhagat (2006). The major argument in these papers was that instead of introducing prioritization techniques, service providers should invest in network capacity that provides sufficient capacity to all end users. Customers are not willing to pay for improved quality (or increased quantity) unless they can observe the change. The reaction in the conventional QoS field was almost non-existent: there are only 16 citations in Google Scholar to these four papers.

As another notable account on the need for QoS that is based on experiences in operational networks is Xiao (2008). The book covers the technical, commercial, and regulatory aspects of QoS in a balanced way. Xiao's proposal for network service providers is to price QoS into their service without introducing different service classes. The main differentiation method in Xiao's proposal is access bit rate. The conclusion made by David Hutchison (2008) in his invited speech at IWQoS seminar was that three classes could be a reasonable solution: premium (includes VoIP), best effort, and less than best effort (for large file transfers). Seth et al. (2007) stressed the need of extensive analysis that integrates technical QoS with non-technical issues drawing on general service quality literature including Parasuraman et al. (1985) and Grönroos (1984). Their framework is very extensive including aspects like security, leadership, and internal communication. Nevertheless, they assume that the service provider shall fulfill QoS guarantees given by the customers.

As Figure 3 shows, the abbreviation QoE started to gain momentum in 2007. QoE was a reaction to the dominant technology-centric attitude in QoS research. The standpoint was clearly explained in the summary of the first QoE seminar in Dagstuhl "From Quality of Service to Quality of Experience" organized in May 2009 (https://www.dagstuhl.de/09192). Similar arguments were presented in Jain (2004) and Kilkki (2008). As shown in Figure 3, maybe 10 or even 20 percent of QoS research was substituted by QoE research.

In addition, net neutrality discussion started during this period, see, e.g., Xiao (2008, Chapter 7), Odlyzko (2009), and Schwartz et al. (2009) and references thereof. Many net neutrality papers also discussed the question of why the Internet lacks service differentiation. One reason for the interest in QoS was the rise of peer to peer (P2P) applications, like BitTorrent. The problems emerged with P2P applications were a kind of antithesis to the connection-oriented model depicted in Figure 1: other users should be protected from the inordinate use of resources required by P2P applications. One may, of course, ask: when do applications' requirements change from something to be satisfied to something to be suppressed? Because of the extent of the question, containing legal, technical, and business aspects, net neutrality papers often provide a broad set of references including historical examples, economic literature, and even some rarely cited, critical QoS papers. In contrast, net neutrality papers are seldom cited by technical QoS papers.

### Back to cellular (2010 – 2017)

As to the most popular QoS topics, a transition from Wi-Fi back to cellular technologies started around 2010 (see Table 3). At the same time, QoS research started to move from the United States to Asia (see Table 4). This shift is not QoS-specific but more general: the importance of Asian universities and private companies in the high-tech area has risen quickly during the last decade (see Baldwin 2016 and Kwon & Kwon 2017).

The 3GPP QoS paradigm explained by Ekström (2009) is based on an aggregate level unit called a bearer with specific aggregate capacity and packet level performance parameters. Each bearer is mapped to one of the standardized QoS classes. The objective of the QoS class system is to ensure that all applications mapped to one QoS class receive similar QoS independent of the network operator and the network equipment vendor. In practice, the QoS class identifier is a reference point for node-specific parameters that, in turn, define connection and packet level treatments, such as scheduling weights, admission thresholds, and packet-level priority. In addition, the 3GPP standards define allocation and retention priorities for the bearers. In principle, the mobile network could apply any treatment to different traffic flows due to the generality of the standard. Nevertheless, the dominant mindset in 3GPP standards is connection-oriented based on applications' requirements.

As to the standardization of QoS schemes and mechanisms, the changes have been minimal during the last ten years. In 3GPP standards, the QoS system with four classes has remained the same although some parameter values were changed, for instance, the maximum bit rates were raised from 2 Mbit/s to 10 Gbit/s. In IEEE standards, the QoS logic and parameters have not been changed. In general, the most notable effort in the field of QoS during the last years has been the development of mappings between the three main QoS systems, see e.g., GSMA (2013, Table 5), IEEE (2016, Table 10-1 and Table R-1), and Kaippallimalil et al. (2015, Table 3).

Papers clearly critical towards QoS are even more difficult to find during this period than in the previous two periods. Perhaps all the relevant arguments were already made without any considerable effect on QoS research. Nevertheless, some valuable summation papers were published. Geoff Huston (2012) argued that the stumbling blocks of QoS are both engineering and economic in nature. The service-quality mechanism chosen must be deployed across all networks along the end-to-end paths of a traffic stream. In the public Internet, this is very difficult due to the array of diverse companies and network entities over any given path. Meddeb (2010) argued that QoS should benefit all users, not only those that can pay for better service. Thus, solving the QoS dilemma requires sociological, cultural, historical, and perhaps even philosophical studies. Similarly, Reichl (2013) emphasized the need for a broader understanding between technical, sociological and economic fields, because any analysis tends to remain impractical if not mapped back to the underlying technology.

As to QoE research, the most active community has organized a series of QoE seminars in Dagstuhl in Wadern, Germany from 2009 to 2016 (see Fiedler 2018). The Qualinet project has also studied various aspects of QoE (Qualinet 2018, Möller & Raake 2014). Those activities have produced numerous articles about the relationship between QoS and QoE. The separation between human (subscriber) and technical (other levels) viewpoints is well-known and relatively clear; practically all QoE papers mention this as an essential aspect. Many QoE papers (e.g., Zhang & Ansari 2011 and Hoßfeld et al. 2018) use three separate levels: user, application, and network. Reichl et al. (2015) use three separate perspectives: system, user, and context. Fiedler et al. (2010) distinguish between a user-centric view (ITU) and a network-centric view (IETF) and argue that the task of QoE research is to make a connection between these views. Brooks & Hestnes (2010) classify stakeholders into three categories: technically oriented, customer oriented, and management. Similarly, Laghari & Connelly (2012) stress the need for multidisciplinary QoE analysis embracing technical, business, and human aspects. Stankiewicz et al. (2011) add legal challenges, e.g., net neutrality, to the overall picture of QoE. In general, although the concept of QoE has extended the scope of traditional QoS studies, it has not been able to solve the big QoS challenge: How to make QoS useful in real networks.

Finally, net neutrality is closely related to the use of QoS in public networks. The discussion started in the previous period and continued actively during this period, see, e.g., Jordan & Ghosh (2010), Statovci-Halimi & Franzl (2013), Schewick (2015), claffy & Clark (2016), and Choi et al. (2018). The net neutrality dilemma is in a sense contrary to the QoS dilemma: promotion of net neutrality emerged as a reaction to network operators' plans to deploy QoS in real networks, not due to the lack of QoS. Thus, even if technical and business challenges could be solved, regulation may limit the deployment of QoS schemes. Even the threat of regulation diminishes the incentives for operators to publicly offer different levels of QoS based on the differences in the willingness to pay or some other business motivation. However, many net neutrality papers are useful to help understand the overall QoS dilemma both from operator and regulator viewpoints.

During this period, we observe a slight increase in citations to the general service quality literature. For instance, Raake & Egger (2014) refers to Parasuraman (1985) and Daniel Kahneman's studies (2003) when discussing the meaning of key terms like quality, utility, and experience. Aziza et al. (2015) is another example that combines general quality literature and QoE literature. These types of citations are still rare in networking QoS or QoE literature, but somewhat more common in closely related fields. For instance, de Souza & Dantas (2015) discuss QoS and QoE in the context of distributed databases. They provide an exceptionally wide base of citations that embrace general service quality literature (e.g., Parasuraman 1985 and Grönroos 1984), network QoE literature (e.g., Bouch et al. 2000, Brooks & Hestnes 2010, and Möller & Raake 2014) and technical QoS literature (e.g., IntServ specification by Shenker et al. 1997).

In summary, there was no change in the prevalent opinion about the meaning and purpose of QoS in the context of communications networks and services. For instance, BITAG (2015) summarizes the benefits of service differentiation as follows: "When differentiated treatment is applied with an awareness of the requirements for different types of traffic, it becomes possible to create a benefit without an offsetting loss." This formulation still closely resembles the connection-oriented process shown in Figure 1.

All the standards remained practically unchanged while the real usage of QoS mechanisms remained limited. It is hard to find any systematic analysis of the use of QoS mechanisms in real networks. Fernández-Segovia et al. (2015) is a rare example in which the researchers have studied the usage of QoS classes in mobile networks. Their results show that only three QoS classes were configured by the network operator: QCI 1 (VoLTE or voice), QCI 5 (multimedia) and QCI 8 (best effort data). Moreover, the share of multimedia was only 0.02% of total traffic while the shares of voice and best effort data were 2.7% and 97.3%, respectively. As to WLAN technologies, Nahrstedt (2012, p. 8) notes that although the 802.11e standard has supported QoS in WLANs, it is not widely deployed in real networks. When Gargoyle (2017) discusses the use of QoS in WLAN routers in their Wiki page, they give as an example a situation in which a student shares his WLAN with his roommate that heavily uses BitTorrent applications. QoS features in WLAN could be used to handle that kind of situation in a fair manner. Finally, they remark "QoS is perhaps the only time in your life you get to decide what is fair."

## QoS as a social phenomenon

All prioritization parameters, rules, and mechanisms rest on the assumption that it would be fair to give better service to some users than others. It is always possible to find some reason to provide better service for a user using an application. For instance, the user can be an important person, she may pay more than others, the application can be important, the application may have special requirements as to delay, reliability or throughput, the session could have been already started and should not be interrupted, or maybe others have already spent too many resources. Anyone can pick one of these reasons, then consider the situation from the viewpoint of the specific user and, finally, conclude that any tool that can improve the service perceived by the specific user is desirable and fair. This kind of reasoning has a strong persuasive power. However, when the process is repeated in the case of every identified reason, the outcome will inevitably be extremely complex with numerous priorities and service classes.

Thus, QoS is, on the one hand, an issue easy to comprehend without any special education, and on the other hand, an extremely complex matter with various conflicting demands and non-linear interactions. Everyone can have seemingly reasonable opinions, while almost no one can understand how all the possible actions described in Table 2 interact with each other and with the feedback mechanisms out of the control of the network. Therefore, any QoS community faces the dilemma between the ease of expressing opinions about QoS and the impossibility of making a comprehensive analysis of a complex QoS system.

In order to be a social system, a QoS community has to define its mission and the principles of communication within the system, for instance, to design and analyze QoS schemes for communication networks. One way to consider the communication occurring inside a social system is to apply Niklas Luhmann's concept of *code of the system* (Luhmann 1995). For instance, in the field of mass media, the code is the distinction between information and non-information (Luhmann 2000). In the case of QoS, the code seems to be the distinction between quality requirements satisfied and not satisfied (as illustrated in Figure 1). There are many benefits of this way of communicating about QoS issues, like easiness to understand, a clear linkage to the fairness considerations creating strong feelings, the suitability for communication with external actors, and the possibility to make a mathematical analysis. The code leaves open the question of who or what is stating the requirements, which is both an advantage as it offers lots of freedom and a source of confusion.

However, only a few people identify themselves professionally as members of a QoS community. Instead, most of the researchers working in the QoS field are primarily members of their working place, be it a network operator, an equipment vendor, a university, or another research organization. PhD students working with QoS topics usually have colleagues working with other topics relating to networking technologies, mathematics, economics, or even psychology (QoE) or law (net neutrality). Those fields have their own codes, rules, and missions.

No global QoS community has emerged to consider the fundamental issues of QoS in communication networks. There is no QoS journal. RIPQOS was organized only once. Although IWQoS seminars have taken place regularly since 1993, IWQoS has remained mostly as a forum to present technical QoS papers. The difficulty in finding the slides of many fascinating keynote speeches given at IWQoS seminars is an illustrative fact. In one of those, Henning Schulzrinne (2010) noted that QoS is attractive because quality sounds good and because it allows both sophisticated math and lab prototypes. In another one, Scott Shenker (2011) noticed that QoS is now mostly an intradomain management issue including traffic engineering to avoid congestion, isolation, and static service classes. It is hard to create a fascinating mission for QoS research based on these observations.

During the last 30 years, several international QoS research projects have brought together like-minded researchers. However, most of the projects were either mathematically oriented or tried to serve the business needs of equipment vendors and network operators. Neither of these viewpoints provides a solid basis for considering the fundamental issues and challenges of QoS in general. Mathematics does not consider the question of whether QoS is needed at all. Similarly, researchers participating in big projects do not have any incentive to question the fundamental assumptions of the project, e.g., the urgent need for QoS.

Thus, the answer to the question "Is there a global community for QoS research?" is no. The conclusion is clear if we examine the most cited QoS papers discussed in the History section: there is a huge variation in the publishing forums while only three out of the 270 most cited papers were presented in an IWQoS workshop.

Should there be a true global QoS community? In principle, the community could allow QoS researchers to be more critical towards the established conventions than what is reasonable in the current situation. However, that kind of task is against the basic principles of social systems, because a social system must promote its own continuation, and critical discourse does not serve that purpose, at least in the short term. The limited effect of the critical keynote speeches at IWQoS workshops on later QoS papers illustrates this phenomenon.

Standardization bodies have intentionally left open how the standardized QoS mechanisms should be used in real networks; instead, they just provide tools to fill Table 2. When some guidelines are offered, they are typically limited to mapping of applications to traffic classes, PHBs, and priorities (e.g., GSMA 2013, Eder et al. 2002, and Annex R in IEEE 2016). The standards do not challenge the assumption about the significance of an application's requirements—which is natural because of the neutral role of standardization bodies. What is needed is workable guidelines that instruct network operators on how high level (business or other) goals could be achieved by exploiting the available QoS mechanisms. One of the few guides is provided by Snir et al. (2003, RFC 3644). The guide also demonstrates how difficult it is to convert all the needs and wishes of human beings, operators' business needs, and requirements and properties of various applications to reasonable actions on the technical levels of packets and flows. No simple or even complex solution can satisfy everyone.

From the viewpoint of social systems, the problem is that operational staff is effectively separated from QoS research and development and business development as well. Still, one of the most critical obstacles to QoS deployment is operational challenges (Mackay & Edwards 2013). It seems obvious that any unnecessary complexity in the network operations and management shall be avoided because of the additional expenses and the additional risk of errors. As an example from an adjacent field, configuration problems cause 10% of service outages while load causes 9% of outages in cloud services, (Gunawi et al. 2016). Although the context is different, the message is clear: while additional QoS mechanisms may alleviate some of the problems related to excessive load, they also are likely to increase configuration errors—to err is human.

Errors are made not only in technical details but also in high-level decisions. Even if a technology is dominant and it obviously satisfies many needs, as the Internet does, this does not automatically indicate that the technology serves the genuine needs of society in an optimal way. One of the tendencies of technical literature is to describe technical systems as a collection of facts of the template "in reality, this requirement is satisfied by this technology in this way" without considering why certain technology has become dominant or what could be the other options. Fortunately, there are valuable exceptions that provide more general consideration like John Day (2008) and many critical papers and presentations (see, e.g., Odlyzko 2003, Schulzrinne 2010, Huston 2012, and Reichl 2015).

Users do not care about anything happening inside the network, maybe not even about QoE, but only about service reliability (Schulzrinne 2010). If this statement is strictly true, there is not much room and need for QoS mechanisms. Why, then, have networking engineers put so much effort to develop intricate QoS systems? Is this simply an error that had led to a huge waste of human effort? We would rather argue that the result is caused by a system-level error. A system level error cannot be remedied by a few individuals, even less by a single paper.

A paper may propose a technical solution to be discussed and developed by a larger community. A paper may even outline a solution to the system level challenges. However, if the change in the way of thinking is large, a paradigm change may occur in the Kuhnian sense (see Kuhn 1970). Perhaps the best example of a paradigm change in the context of communications networks occurred when the Internet and Web disturbed the great plans of telecom operators. The change created a need for co-operation between several network technologies with different origins and principles. The confusion about the basic principles of QoS is a part of that struggle to define the intrinsic nature of communications services.

First, we shall ask whether the current system code (requirements satisfied/not satisfied) is optimal or at least satisfactory. If the answer is yes, then there are two main questions: Who shall define the requirements? What is the relative importance of the stated requirements? If the answer to the code question is no, then community shall define a new code to communicate about QoS. A change in the system code could trigger essential changes not only in the design of QoS schemes but also in QoS standards. Any significant change in QoS standards appears unrealistic. Thus, any realistic novel QoS scheme needs to utilize the QoS parameters defined in the current standards. Instead of driving a revolution on the technical layer, we shall promote common effort to change the way we think about QoS.

## The state of QoS standards

Let us start the design of the new QoS program by inspecting the current QoS standards and the proposed QoS mappings between the standards. The unambiguous agreement is that to make QoS widely useful, different types of networks and different network operators must systematically cooperate with each other. Various mappings between different standards have been proposed, see, e.g., Table 5 in GSMA (2013), Tables 10-1 and R_1 in IEEE (2016), and Table 3 in Kaippallimalil et al. (2015). The proposed mappings are similar, but there are some minor discrepancies. Table 4 provides a kind of summary of those mappings. Although some authors (e.g., Szigeti et al. 2018, RFC 8325) introduce even more complicated mappings, we have not included them in Table 5.

Table 5. A mapping between 3GPP, IETF, and IEEE QoS schemes. Green colored rows indicate the four primary QoS classes.

| 3GPP QoS Traffic class | THP | QCI | DiffServ PHB | 802.11 EDCA AC | 802.1D UP |
|---|---|---|---|---|---|
|  | N/A | 1 | EF | AC_VO | 7 |
| Conversational | N/A | 2 | EF | AC_VO | 6 |
|  | N/A | 3 | EF | AC_VI | 5 |
| Streaming | N/A | 4 | AF41 | AC_VI | 4 |
| Interactive | 1 | 5 | AF31 | AC_BE | 3 |
|  | 2 | 6 | AF21 | AC_BE | 3 |
|  | 3 | 7 | AF11 | AC_BE | 0 |
| Background | N/A | 8 | BE | AC_BK | 2 |
|  | N/A | 9 | BE | AC_BK | 1 |

Is this what we are proposing as a new QoS solution? No, because there is no clear understanding what all these classes, indicators, categories, and priorities mean (or should mean) in reality. They are abstract concepts with several possible interpretations. For example, the term class simply implies some differences between the properties of the classes and some similarities in the needs of the users or applications in the same class. PHB is an even vaguer concept. Even priority, which appears clearer, can mean different things. In Table 2, priority is mentioned in several places. However, to have a concrete effect a priority from the *Change the status* column in Table 2 must be translated to actions shown in the other columns in Table 2: serving now or later, rejecting or changing the allowed bit rate. The aim of high priority can be to offer short delay, small drop probability, or high capacity. We tend to assess all three aspects on the same emotional scale that could be called importance. We can consider all kinds of use cases, applications, users, and customer segments, compare them to other entities and, finally, locate each of them on a scalar importance scale. It might even be possible to find a rough consensus about the ordering.

The result is acceptable from the viewpoint of many social systems including QoS research and standardization, but very problematic from the viewpoint of those who need to operate and manage a network and interoperations between different networks.

However, is there any other option except to put almost everything to the best effort class? In addition to best effort, a few cases with special needs could be handled separately. This is, anyway, how most networks are nowadays managed. Voice is the most important special case. Telephone service was a great business for over 100 years. Nowadays, there are many other valuable applications like video streaming, video conference, web browsing, online games, file transfer, and e-mail. Because voice is entitled to a special service and because these applications create similar feelings of importance as phone calls, we argue that they also are entitled to their own special service.

Why do network operators offer phone service but very few other application-specific services? Is the reason that voice applications are designed in a way that they require low delay or bit rate guarantees and the network can satisfy those requirements only by a special voice service?

Not really, because the real requirements are human-centric, not technical. Voice is a special service because 1) speech is a natural way of communicating among all people 2) during a voice call, two persons direct their (highly valuable) attention to the dialog, 3) a dialog starts to be disturbed if the round-trip delay exceeds 250 ms, 4) abrupt interruptions during a dialog are very annoying, and 5) the amount of transmitted information during a dialog is relatively small. A voice application is not valuable; human communication is valuable. A voice application does not require anything without two human beings.

In practice, best effort service is appropriate for web browsing, file transfer, and e-mail without any special arrangements. Video streaming, video conference, and online gaming are more complex cases. Compared to voice, video streaming has the following properties: 1) the data volumes are much larger, 2) the level of attention and the value per minute are somewhat lower, 3) delay requirement is less strict (due to the possibility of caching data), 4) interruptions are somewhat less annoying (note that ads are often tolerated during videos but not during phone calls). As a result, the value of video streaming <u>per used resource</u> is much smaller than that of a phone call. Moreover, it is unclear whether the value of a streaming video per used resource is higher than that of other typical uses of best effort service. Odlyzko (2012) offers an insightful analysis on the value of information—this aspect deserves more attention also in the QoS context.

It is hard for a human mind to make multiplications and divisions with large numbers. Feelings are not changed when the scale is changed, for instance, from 20 kbit/s to 2 Mbit/s. These are just numbers that do not create any feelings. Thus, it does not seem reasonable to rely directly on feelings when assessing how different applications should be treated in a communications network. Placing applications in order of importance based on intuition and feelings is a futile, if not harmful, exercise.

For instance, in some online shooting games, a player may get a competitive edge due to a shorter network delay. From the network operator viewpoint, only if a player is willing to pay more than a regular customer, does it make sense for the operator to provide better service for the player. However, it is not necessarily in the interest of the gaming community to encourage players to pay more for network service to achieve a comparative advantage.

In brief, any classification based on the feelings different applications create in our minds is infeasible. In a commercial context, an analysis of the business potential is a better approach. Unfortunately, there is a strong tendency to end up with the same list of applications and their requirements with which we started our discussion. Thus, we need an alternative way of thinking.

### Incentive-based QoS

As discussed in the History section, the main criticisms of the current QoS systems are: the overall complexity of QoS systems, the additional cost of operations and management, and the lack of viable business models. Furthermore, it has been noted that both end-users and application developers dislike all additional QoS features that do not offer obvious advantages for them. Finally, only during periods of mild to moderate congestion can QoS improve the value of network service; in periods without congestion, there is no need to do anything while during severe congestion, the overall service quality is inevitably very low. Thus, all critical authors promote the use of the best effort model and overprovisioning as the preferred solution.

Any novel QoS systems must give credible solutions to the identified problems. In this section, we outline the principles of a QoS scheme based on incentives instead of requirements. In the incentive-based QoS scheme, a network operator interacts with users and with adjacent network operators by means of pricing and traffic control rules that create incentives for external actors to use the network resources appropriately. Appropriate means here that the outcome is beneficial for both the network operator and the users. Thus, this scheme is not based on the requirements something or someone provides to the network, but a structure of incentives provided by the network to users and applications. As an indication of the difference in the viewpoint, the term incentive does not appear in any of the first three sentences with QoS in the 270 most cited papers, while requirement appears 221 times in 131 papers.

Incentives work on two layers, pricing and traffic control. In terms of pricing, in principle, a customer paying more may get preferential treatment to the access network. This principle is used when the price is based on the (maximum) access bit rate. In addition to the access bit rate, the network operator may define the maximum amount of data that the user is allowed to download from (or transmit to) the network over a period, typically a month. If the limit is exceeded, either the access rate is throttled or the customer has to pay more. This creates an incentive for the customer to use the network resources cautiously even when a pure best effort service is used inside the network. If we use the terminology of Table 1, the viewpoint is subscriber.

Alternatively, traffic control and the design of such control is the most important property of the incentive-based QoS. However, the operator should use traffic control or QoS mechanisms only if there are good reasons that justify the additional operational burden. A reasonable justification for QoS is a situation in which the network remains congested because a minority of users permanently fill the network with excessive traffic. Without congestion, there is hardly any reason to use QoS mechanisms. If the traffic load is evenly distributed amongst the users, the preferred solution is to increase the network capacity. In this case, the operator may rely on the ability of TCP to divide the available capacity evenly between all active users.

What should a network operator do if it is too costly to upgrade the network so that a best effort service model provides good enough performance for all users? We answer that the network operator should provide incentives for users to adjust their traffic patterns in a way that 1) effectively alleviates the congestion situation, 2) is acceptable for all users, and 3) supports operator's business goals. In an optimal situation, the reaction is automatic and happens without any conscious effort by the user. However, it might be useful to have an option in which the user can declare her preferences with regard to the importance of different applications or traffic flows.

The idea of incentives is to induce a desirable reaction to a specific situation. The primary idea of incentive-based QoS is that QoS rules make it possible for the device or application to make reasonable decisions automatically. Also, we can presume that TCP is widely used to adjust the bit rates used by many applications.

What then is feasible, if TCP takes care of all moderate congestion situations and if we exclude dynamic pricing schemes that tend to annoy users? Two issues may create the need for additional incentive-based QoS mechanisms: 1) Situations where the importance cannot be deduced from information about the application or the user and 2) situations with considerable differences in the relative importance between delay and bit rate.

In the first case, the incentive shall be built in a way that makes it possible for a user to get a larger share of capacity than normal for a limited period at the expense of a reasonable cost (e.g., lower capacity later). In this approach, the network must keep track of individual users and their usage over time—this sounds reasonable in cellular networks with identified subscribers, but problematic in networks without reliable user identification. Instead of subscriber, the unit could be device in WLAN networks or IP address in the Internet. However, the fundamental problem in this approach is that operators likely ignore any momentary (higher than best effort) priority that has authorized by another operator.

In the second case, we assume that some applications are particularly sensitive to delay and jitter (e.g., some online games) while other applications are insensitive to delay but can use as much transmission capacity as possible (e.g., file transfer). If a network offers three delay classes, the highest class could provide as low delay and jitter as the networking technology allows, the second class could be suitable for voice, while the third class could provide the delay properties of the current best effort service.

How can we construct incentives in a way that the user or the application is allowed to select the delay class freely? The principle proposed here is that the application trades off between delay and bit rate: higher class (i.e., lower delay) always means a lower share of capacity. If during a congestion situation, a best effort flow receives 100 units of capacity, a flow in voice class may receive 30 units of capacity, and a flow in the highest class, 10 units of capacity. The differences in the relative shares shall be large enough to prevent unnecessary use of higher delay classes. The basic logic of the incentive-based QoS is illustrated in Figure 4. In brief, each application is free to select any of the delay classes and send traffic at any bit rate, but those decisions affect the likelihood that traffic (or packets) will be dropped inside the network. The user's role is just to decide whether the experience is good enough to justify the consumed time. Depending on the business model, the user may also decide to upgrade (or downgrade) her customer contract to get better pricier service (or cheaper worse service).

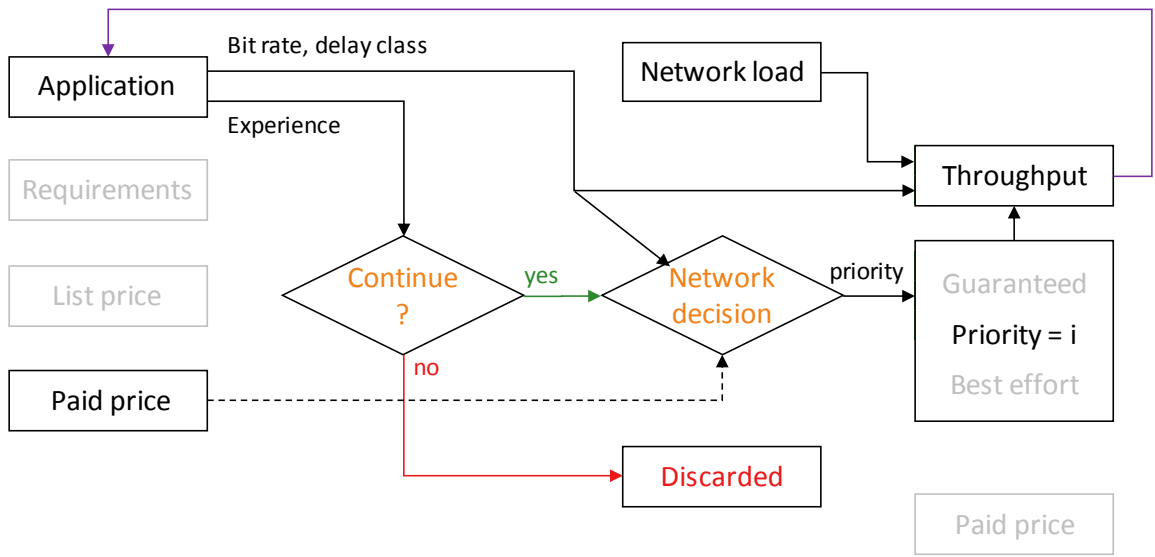

Figure 4. An outline of the service process based on the incentive-based QoS.

As to the technical implementation, there are already four classes available in all major technologies as Table 5 shows (columns: traffic class, PHB, and EDCA AC and rows: conversational, streaming, interactive, and background), though the names are not necessarily the best possible in the context of incentive-based QoS. The other technical issue is how to enforce the desired differences in shares for flows selecting different delay classes. The simplest solution would be limit the traffic in each delay class in a similar way as the overall access rates for a subscriber is limited.

A more dynamic and sophisticated (i.e., complex) solution would be to use the principles of Simple Integrated Media Access (SIMA) described in Kilkki (1999, chapter 7.5) and covered by two US Patents (Kilkki 2000a, 2000b). Note that because the filing date of the patents was March 20, 1997, the patents are not anymore in force. A SIMA-type solution means a sharp separation between delay class and packet level priority. In SIMA, priorities are used to divide the available capacity in a controlled way. The priority of a packet is calculated based on the momentary bit rate: the higher the bit rate, the lower the priority, and the higher the probability of packet discarding. This creates a strong incentive for the user (or application) to adjust whenever there is congestion inside the network. Practical use of SIMA requires two (or maybe three) delay classes and a sufficient number of bits to define the drop priorities for IP packets (or some similar units). Three bits, which means eight priorities, would be enough to cope with most of the congestion situations. In this kind of incentive-based scheme, it is not reasonable to define a mapping between service classes and priorities in a way similar to Table 5, because (delay) classes and (drop) priorities are orthogonal scales.

However, SIMA contains similar problems as any other complicated QoS scheme. A QoS scheme creates a reliable end-to-end service only if all network domains of the path are using similar QoS schemes. Even if the schemes are similar, there might be a need to design specific rules and markings for solving interconnection problems in a similar way as described by Geib & Black (2017, RFC 8100). Moreover, it might be possible to create incentives between the operators in a similar way as between a network operator and the users of the network.

It should be stressed that incentive-based QoS can be based on constructions other than SIMA. What we are promoting here is the fundamental principles of incentive-based QoS, not any specific realization of incentives. The main objective of QoE studies in the context of incentive-based QoS would be to analyze the trade-offs between delay-class, bit rate, and reliability (measured through packet loss ratio). Better delay class means a lower bit rate when reliability is kept constant. Lower bit rate means higher reliability when delay class is kept constant. The result of the analysis depends on the properties of the application and the preferences of users.

As to net neutrality, the critical issue is how the incentives are constructed. Incentives can and should be constructed in a way are they are independent of the application and the user. Each application and each user should be entitled to react to the given incentives as they consider appropriate. Still, the operator must design the incentives in a way that treats all users and other agents in a fair manner.

Finally, when an effective incentive-based QoS is in operational use, the main task for the network operator is to optimize the network capacity in different parts of the network. In principle, this is the same task as in the case of a best effort network, but now with automatically adjusted quality differentiation.

But does this incentive-based QoS solve the problems identified at the beginning of this section? The approach is certainly simpler to operate and manage than the connection-oriented QoS model illustrated in Figure 1 because in the incentive-based model the network does not directly care about applications and their requirements. Compared to the fuzzy construction of traffic classes and priorities (outlined in Table 5), the proposed implementation of incentive-based QoS gives exact meaning for traffic classes (delay) and priorities (importance). This exactness makes operational actions more straightforward and alleviates the interoperability problems between network operators. As to the business model challenge, incentive-based QoS allows the use of both a single price class or multiple price classes. The possible price differentiation affects directly the relative amount of resources obtained by different customers; the main change compared to the current business model is that price class not only defines the access bit rate but also how resources are divided in the core network. One of the main advantages of the incentive-based QoS is that, when properly designed, it works automatically during severe congestion situations without changing the current best-effort model during normal conditions.

## Discussion

For some of us, QoS has been an important topic during our professional career. Tens of thousands of QoS papers have written with a huge effort and cost. However, when looked from a broader perspective, QoS has not been a particularly important topic in the area of communications. For instance, when John Day (2008) discussed the general problems in networking, he spent only about one page (out of 383 pages) for discussing QoS issues. He observed that only a few networking protocols "have paid more than lip services to doing anything about [QoS] (to some extent with good reason)." As another example, quality is not among the top 50 keywords in the papers published in Telecommunications Policy (Kwon & Kwon, 2017). As a third example, in Kevin Allocca's (2018) new book about how YouTube videos are changing the world he does not mention QoS at all. However, videos are often given as the major justification for the use of QoS mechanisms.

It seems that some of us long for powerful QoS as it creates positive feelings. There might be different reasons for the longing, but hardly ever the reason is a concrete event in which QoS has saved the day. The reverse is much more likely; you were frustrated due to poor service and expected the network operator to do something to improve the quality of service. These personal feelings are easily mixed with analytical considerations of such a complex issue as QoS.

A proper QoS scheme (as well as analysis) should be based on the value of the service for the users and the service provider rather than the applications technical requirements. The value for the users is known only to the user and cannot be inferred from the application. Thus, in the case of insufficient network resources, the user shall, together with automatic rules built in the application, decide how to react to the incentives designed by the network operator. Besides, we need to remember that the feelings created at a service instant (say, a video streaming session, a voice call or online game) do not depend on the amount of resources utilized at that service instance. This cost part of the analysis must be carried out separately. The cost, particularly in cases with huge differences in required transmission capacity, is the main justification for incentive-based QoS.

Another reason for using QoS methods is to prepare for unpredictable situations where transmission capacity is severely reduced, or demand is increased due to an emergency or a network attack. In this kind of situation, the network must react automatically without complex rules and resource allocation policies. As Katz et al. (2005) expresses it, "Rather, we need survival policies when the network is under stress." In the incentive-based QoS, the problem of survival policies can be automatically solved if there are enough priority levels to provide sufficient dynamics in the incentive system.

Therefore, after a long search for the true need for quality of service, we are ready to summarize our findings:

1. In normal conditions, the default solution is to use best effort service and overprovisioning without quality differentiation. Only if the network cannot provide sufficient capacity to satisfy all reasonable needs of users with an acceptable cost, should the use of quality differentiation or QoS be considered.
2. When quality differentiation is truly needed, design a QoS scheme that provides appropriate incentives for users to adjust their traffic patterns in a manner that is beneficial for both the network operator and the users.
3. Leave the design of algorithms for reacting to network incentives for application developers.
4. Design the scheme in a way that the network work also in exceptional situations in a reasonable way.
5. Build a common agreement with other network operators and other key players in the field to develop and use compatible incentive-based QoS schemes.

The good news is that all this can be achieved on a technical level using marking on the packet (and perhaps aggregate) level and some traffic control on the device or subscriber level without making any essential change to the current network standards. The bad news is that this would require a paradigm change from requirements to incentives. In the terminology of Niklas Luhmann, that would mean a change in the code of the system from "quality requirements satisfied/not satisfied" to "good incentive/bad incentive." A paradigm change is hard to achieve, but the minimal use of QoS mechanisms in real networks motivates stakeholders to reshape the whole QoS field.

## Future

This is a long paper with far-reaching ideas. As the literature analysis in this paper has also shown, the most cited papers seldom challenge the dominant way of thinking. In contrast, the fate of the most critical papers is often indifference. For instance, Google Scholar finds one citation to Bhagat (2006), two to Huston (2012), and two to Reichl (2013).

Quality itself is an elusive term and does not convey all important aspects of a paper or a service. The quality of a scientific paper is typically measured through a peer-review process leading to a recommendation to either accept or reject a submitted paper. This process resembles the Mean Opinion Score used in QoE studies and Net Promoter Score used in customer relationship studies. Any opinion expressed in the submitted paper that the reviewer intuitively disagrees with generates negative feelings that are likely to result in a negative assessment even when the reviewer appreciates other parts of the paper. Generally, bad feelings are much stronger than good feelings as thoroughly discussed by Baumeister et al. (2001). This innate property of the human mind may lead to papers that minimize the likelihood of disagreements.

When the authors, nevertheless, decide to take the risk and challenge the prevalent opinion, they must decide whether to go through the normal review process. Because of the nature of this paper, we decided to publish a first version without any peer review. Instead, we provide incentives for the readers to utilize the paper as efficiently as possible.

You can refer to this version, and you can give direct comments to kalevi.kilkki@aalto.fi. Based on the comments we will prepare a new version of the paper in which we use all valuable comments and questions as additional references (either anonymously or by name depending on your preference). You may also propose to add other references. Our long-term aim is to promote a paradigm change in the way we think about QoS. In that endeavor, one paper can never be enough. It takes a community to change a paradigm.

# Appendix. List of abbreviations

| | |
|---|---|
| 3G | Third Generation |
| 3GPP | Third Generation Partnership Program |
| 5G | Fifth Generation |
| AC | Access Category |
| ACM | Association for Computing Machinery |
| AF | Assured Forwarding |
| ARPANET | Advanced Research Projects Agency Network |
| ATM | Asynchronous Transfer Mode |
| BE | Best Effort |
| CCITT | Comite Consultatif International de Telegraphique et Telephonique |
| COST | Committee on Science and Technology |
| DiffServ | Differentiated Services |
| EDCA | Enhanced Distributed Channel Access |
| EF | Expedited Forwarding |
| ETSI | European Telecommunications Standards Institute |
| GPRS | General Packet Radio Service |
| GSM | Global System for Mobile Communications |
| GSMA | GSM Association |
| IEEE | Institute of Electrical and Electronics Engineers |
| IETF | Internet Engineering Task Force |
| IntServ | Integrated Services |
| IP | Internet Protocol |
| ISO | International Organization for Standardization |
| ISP | Internet Service Provider |
| ITU | International Telecommunication Union |
| ITU-T | ITU -Telecommunication Standardization Sector |
| IWQoS | International Symposium on Quality of Service |
| MAC | Media Access Control |
| MB | Megabyte |
| MPLS | Multiprotocol Label Switching |
| P2P | Peer to Peer |
| PHB | Per Hop Behavior |
| PhD | Philosophiae Doctor |
| QCI | QoS Class Identifier |
| QoE | Quality of Experience |
| QoS | Quality of Service |
| RFC | Request For Comment |
| RIPQOS | Revisiting IP QoS (workshop) |
| RSVP | Resource Reservation Protocol |
| SIM | Subscriber Identity Module |
| SIMA | Simple Integrated Media Access |
| SLA | Service Level Agreement |
| TCP | Transmission Control Protocol |
| THP | Traffic Handling Priority |
| TOS | Type of Service |
| UMTS | Universal Mobile Telecommunications System |
| UP | User Priority |
| VC | Virtual Connection |
| VoIP | Voice over Internet Protocol |
| VoLTE | Voice over Long Term Evolution |
| VP | Virtual Path |
| VPN | Virtual Private Network |
| WiMAX | Worldwide Interoperability for Microwave Access |
| WLAN | Wireless Local Area Network |